# HOW A PLANTAR PRESSURE-BASED, TONGUE-PLACED TACTILE BIOFEEDBACK MODIFIES POSTURAL CONTROL MECHANISMS DURING QUIET STANDING

Nicolas VUILLERME, Nicolas PINSAULT, Olivier CHENU, Matthieu BOISGONTIER,

Jacques DEMONGEOT and Yohan PAYAN

Laboratoire TIMC-IMAG, UMR CNRS 5525, La Tronche, France

Address for correspondence:

Nicolas VUILLERME

Laboratoire TIMC-IMAG, UMR CNRS 5525

Faculté de Médecine

38706 La Tronche cédex

France.

Tel: (33) (0) 4 76 63 74 86

Fax: (33) (0) 4 76 51 86 67

Email: nicolas.vuillerme@imag.fr

**Abstract**

The purpose of the present study was to determine the effects of a plantar pressure-based, tongue-placed tactile biofeedback on postural control mechanisms during quiet standing. To this aim, sixteen young healthy adults were asked to stand as immobile as possible with their eyes closed in two conditions of No-biofeedback and Biofeedback. Centre of foot pressure (CoP) displacements, recorded using a force platform, were used to compute the horizontal displacements of the vertical projection the centre of gravity ($CoG_h$) and those of the difference between the CoP and the vertical projection of the CoG ($CoP-CoG_v$).

Analysis of the $CoP-CoG_v$ displacements showed larger root mean square (RMS) and mean power frequencies (MPF) in the Biofeedback than No-biofeedback condition. A stabilogram-diffusion analysis further showed a concomitant increased spatial and reduced temporal transition point co-ordinates at which the corrective processes were initiated and an increased persistent behaviour of the $CoP-CoG_v$ displacements over the short-term region.

Analysis of the $CoG_h$ displacements showed decreased RMS and increased MPF in the Biofeedback relative to the No-biofeedback condition. A stabilogram-diffusion analysis further indicated that these effects mainly stem from reduced spatio-temporal transition point co-ordinates at which the corrective process involving $CoG_h$ displacements is initiated and an increased anti-persistent behaviour of the $CoG_h$ displacements over the long-term region.

Altogether, the present findings suggest that the main way the plantar pressure-based, tongue-placed tactile biofeedback improves postural control during quiet standing is via both a reduction of the correction thresholds and an increased efficiency of the corrective mechanism involving the $CoG_h$ displacements.

**Introduction**

We recently developed an original biofeedback system for improving balance whose underlying principle consists in supplying the user with supplementary sensory information related to foot sole pressure distribution through a tongue-placed tactile output device. In a pioneering study, the effectiveness of this system in improving postural control during quiet standing has been established, suggesting that an artificial tongue-placed tactile biofeedback can be efficiently integrated with other sensory cues by the postural control system (Vuillerme et al. 2006a). At this point, however, no information was provided regarding how the central nervous system (CNS) uses this biofeedback information for postural control, or, in other words, how this plantar pressure-based, tongue-placed tactile biofeedback modifies the control mechanisms involved in postural control during quiet standing.

In the above-mentioned investigation (Vuillerme et al. 2006a), the displacements of the centre of foot pressure (CoP), whose successive positions express the displacements of the point of application of the resultant reactive force, were used to characterise postural control. During bipedal quiet standing, however, this CoP assumes two distinct tasks: it counteracts the centre of gravity (CoG) in its falling and makes it remain in a particular zone within the base of support. In other words, the CoP displacements are aimed at facilitating the displacements of the CoG to return to a position more compatible with equilibrium, in the first case, and reducing the displacements of the CoG as much as possible, in the second case (Winter et al. 1996; Rougier and Caron 2000).

From this, it seems pertinent to dissociate the CoP into two elementary components, (1) the horizontal displacement of the vertical projection the centre of gravity ($CoG_h$) and (2) those of the difference between the CoP and the vertical projection of the CoG ($CoP-CoG_v$) (e.g., Rougier and Caron 2000; Rougier and Farenc 2000), presenting specific attributes in postural control. The former, representing the whole body motions, can be considered as the controlled variable during bipedal quiet standing (e.g. Clément et al. 1984; Horstmann and Dietz 1990; Winter et al. 1998), whereas the latter, in addition to demonstrate a certain proportionality with the horizontal acceleration communicated to the $CoG_h$ (Brenière et al. 1987; Gage et al. 2004; Winter et al. 1998), is assumed to express the ankle joint stiffness (Caron et al. 2000; Winter et al. 1998) and to be linked to the level of neuromuscular activity (Rougier et al. 2001). Thus, by decomposing the CoP trajectory into two elementary motions, it is possible to determine to which extent a modification of the global CoP displacements can arise from either a single exaggerated elementary motion or both of them.

Furthermore, applying mathematical concepts such as fractional Brownian motion (fBm) (Mandelbrot and van Ness 1968) to the $CoG_h$ and $CoP-CoG_v$ displacements allows to gain additional insight into the control mechanisms used for controlling these displacements during quiet standing (e.g. Rougier and Caron 2000; Rougier and Farenc 2000). Indeed, through this model and the resort to the so-called "stabilogram-diffusion analysis" (e.g. Collins and De Luca 1993), the temporal organisation of various control mechanisms involved in controlling upright stance can be determined in the sense that two distinct control mechanisms, "persistent" and "anti-persistent" operate in continual succession (e.g. Collins and De Luca 1993). Interestingly, this model allows the determination of the spatiotemporal characteristics of the switch between these successive mechanisms, or, in other words, from when and to what extent the corrective process begins to operate, on average.

Within this context, the purpose of the present experiment was to investigate in more depth the effects of an artificial plantar-based, tongue-placed tactile biofeedback on postural control mechanisms during quiet standing. Analyses of $CoP-CoG_v$ and $CoG_h$ displacements and resort to the fBm framework through the stabilogram-diffusion analysis should indicate to which extent the decreased CoP displacements, recently observed when biofeedback was in use relative to when it was not (Vuillerme et al., 2006a), can be explained by (1) a modification of the respective contributions of CoP-CoG$_v$ and/or CoG$_h$ motions in the global CoP trajectories on the one hand and/or (2) the subjects' ability to control these elementary motions in a more or less precise manner on the other hand.

**Materials and Methods**

Subjects

Sixteen young university student (age: 24.8 ± 3.2 years; body weight: 71.4 ± 15.0 kg; height: 177.3 ± 11.8 cm) participated in the experiment. They gave their informed consent to the experimental procedure as required by the Helsinki declaration (1964) and the local Ethics Committee, and were naive as to the purpose of the experiment. None of the subjects presented any history of motor problem, neurological disease or vestibular impairment.

Task and procedures

Subjects stood barefoot, feet together, their hands hanging at the sides, with their eyes closed. They were asked to sway as little as possible in two No-biofeedback and Biofeedback conditions. The No-biofeedback condition served as a control condition. In the Biofeedback condition, subjects performed the postural task using a plantar pressure-based, tongue-placed tactile biofeedback system. A plantar pressure data acquisition system (FSA Inshoe Foot pressure mapping system, Vista Medical Ltd.), consisting of a pair of insoles instrumented with an array of 8 × 16 pressure sensors per insole (1cm² per sensor, range of measurement: 0-30 PSI), was used. The pressure sensors transduced the magnitude of pressure exerted on each left and right foot sole at each sensor location into the calculation of the positions of the resultant ground reaction force exerted on each left and right foot, referred to as the left and right foot centre of foot pressure, respectively (CoP$_{lf}$ and CoP$_{rf}$). The positions of the resultant CoP were then computed from the left and right foot CoP trajectories through the following relation (Winter et al. 1996):

CoP = CoP$_{lf}$ × R$_{lf}$ / (R$_{lf}$ + R$_{rf}$) + CoP$_{rf}$ × R$_{rf}$ / (R$_{rf}$ + R$_{lf}$),

where R$_{lf}$, R$_{rf}$, CoP$_{lf}$, CoP$_{rf}$ are the vertical reaction forces under the left and the right feet, the positions of the CoP of the left and the right feet, respectively.

CoP data were then fed back in real time to a recently developed tongue-placed tactile output device (Vuillerme et al. 2006a,b,c). This so-called Tongue Display Unit (TDU), initially introduced by Bach-y-Rita et al. (1998), comprises a 2D array (1.5 × 1.5 cm) of 36 electrotactile electrodes each with a 1.4 mm diameter, arranged in a 6 × 6 matrix. The matrix of electrodes, maintained in close and permanent contact with the front part of the tongue dorsum, was connected to an external electronic device triggering the electrical signals that stimulate the tactile receptors of the tongue via a flat cable passing out of the mouth. Note that the TDU was inserted in the oral cavity all over the duration of the experiment, ruling out the possibility the postural improvement observed in the Biofeedback relative to the No-biofeedback condition to be due to mechanical stabilization of the head in space. The underlying principle of our biofeedback system was to supply subjects with supplementary information about the position of the CoP relative to a predetermined adjustable "dead zone" (DZ) through the TDU. In the present experiment, antero-posterior and medio-lateral bounds of the DZ were set as the standard deviation of subject's CoP displacements recorded for 10 s preceding each experimental trial. A simple and intuitive coding scheme for the TDU, consisting in a "threshold-alarm" type of feedback rather that a continuous feedback about ongoing position of the CoP, was then used. (1) When the position of the CoP was determined to be within the DZ, no electrical stimulation was provided in any of the electrodes of the matrix. (2) When the position of the CoP was determined to be outside the DZ, electrical stimulation was provided in distinct zones of the matrix, depending on the position of the CoP relative to the DZ. Specifically, eight different zones located in the front, rear, left, right, front-left, front-right, rear-left, rear-right of the matrix were defined; the activated zone of the matrix corresponded to the position of the CoP relative to the DZ. For instance, in the case that the CoP was located towards the front of the DZ, a stimulation of the anterior zone of the matrix (i.e. stimulation of the front portion of the tongue) was provided. Finally, in the present experiment, the frequency of the stimulation was maintained constant at 50 Hz across participants, ensuring the sensation of a continuous stimulation over the tongue surface. The intensity of the electrical stimulating current was adjusted for each subject, and for each of the front, rear, left, right, front-left, front-right, rear-left, rear-right portions of the tongue, given that the sensitivity to the electrotactile stimulation was reported to vary between individuals (Essick et al. 2003), but also as a function of location on the tongue in a preliminary experiment (Vuillerme et al. 2006b). Several practice runs were performed prior to the test to ensure that subjects had mastered the relationship between the position of the CoP relative to the DZ and lingual stimulations.

A force platform (AMTI model OR6-5-1), which was not a component of the biofeedback system, was used to measure the displacements of the centre of foot pressure (CoP), as a gold-standard system for assessment of balance during quiet standing. Signals from the force platform were sampled at 100 Hz (12 bit A/D conversion) and filtered with a second-order Butterworth filter (10 Hz low-pass cut-off frequency).

Three 30s trials for each experimental condition were performed. The order of presentation of the two experimental conditions was randomized.

Estimation of the $CoG_h$ and $CoP\text{-}CoG_v$ displacements

$CoG_h$ and $CoP\text{-}CoG_v$ motions were determined from the CoP trajectories computed from the force platform. A relationship between the amplitude ratio of the $CoG_h$ and CoP motions ($CoG_h$/CoP) and sway frequencies allowed determining the $CoG_h$ and consequently $CoP\text{-}CoG_v$ motions. Body sways being particularly reduced, standing still can therefore theoretically be modelled as a one-link inverted pendulum (e.g. Winter et al. 1998; Gage et al. 2004), where $CoG_h$ and CoP behave as periodic functions in phase with each other. The method, initially proposed by Brenière (1996) for gait and then extended to standing posture by Caron et al. (1997), is given by the following formula:

$CoG_h/CoP = \Omega_0^2/(\Omega_0^2 + \Omega^2)$,

where $\Omega = 2\pi f$ is the pulsation (rad s$^{-1}$) and $\Omega_0 = [mgh / (I_G + mh^2)]^{1/2}$ (Hz), termed natural body frequency, is a biomechanical constant relative to the anthropometry of the subject (m, g, h, IG: mass of the subject, gravity acceleration, distance from CoG to the ground, and moment of body inertia around the ML or AP axis with respect to the CoG).

From this $CoG_h$/CoP relationship, it is therefore relevant to consider that CoP oscillations operating over too high frequencies would not incur appreciable $CoG_h$ movements. The principle of this model is that the body constitutes a low-pass filter, which would explain the loss in amplitude observed between CoP and $CoG_h$ as the sway frequency increases. The $CoG_h$ estimation consists in multiplying the data, transformed in the frequential domain through a Fast Fourier Transform (FFT), by the above-mentioned filter and recovering to the time domain by processing an inverse FFT (e.g. Berger et al. 2005; Bernard-Demanze et al. 2006; Rougier et al. 2001; Rougier and Caron 2000; Rougier and Farenc 2000).

Data analysis

$CoP\text{-}CoG_v$ and $CoG_h$ displacements were processed through two different analyses.

(1) A frequency-domain analysis, issued from the FFT process, included the calculation of (*i*) the root mean square (RMS) and (*ii*) the mean power frequency (MPF) parameters aimed at characterising the mean spectral decompositions of the sway motions on specific bandwidths (0-0.5 Hz for $CoG_h$ and 0-3 Hz for $CoP\text{-}CoG_v$). These bandwidths were chosen to give these indices the larger sensitivity since the modifications occurring on the frequency spectra intervene generally inside these bounds.

(2) A stabilogram-diffusion analysis (Collins and De Luca 1993) as described initially by Mandelbrot and van Ness (1968) enabled the assessment of the degree to which a trajectory is controlled. This degree is indeed appreciated through the half-slope of a variogram expressing the mean square displacements $<\Delta x^2>$ as a function of increasing time intervals $\Delta t$. A median value of 0.5 for this half-slope, through which the scaling exponent H is computed, indicates a lack of correlation between past and future increments and suggests a complete lack of control. On the other hand, i.e. if H differs from 0.5, positive (H>0.5) or negative (H<0.5) correlations can be inferred, which is indicative of a given part of determinism of the control. Depending on how H is positioned with respect to the median value 0.5, it can be inferred that the trajectory is more or less controlled: the closer the scaling regimes are to 0.5, the lesser the control. In addition, depending on whether H is superior or inferior to the 0.5 threshold, persistent (the point is drifting away) or anti-persistent behaviours (the point retraces its steps) can be revealed, respectively. Thus, for each of the two elementary motions and each ML and AP axis, the stabilogram-diffusion analysis included the calculation of (i) the temporal ($\Delta t$) and spatial ($<\Delta x^2>$) co-ordinates of the transition point and (ii) the two scaling exponents, indexed as short ($H_{sl}$) and long latencies ($H_{ll}$).

Statistical analysis
    Data from both No-biofeedback and Biofeedback conditions were compared through *t*-tests, the first level of significance being set at 0.05.

**Results**
    Before presenting the results, it important to emphasis that no statistical differences have been noticed regarding the mean positioning of the CoP between the two conditions of No-biofeedback and Biofeedback along both the ML and AP axes ($Ps>0.05$).

Frequency-domain analysis
    Analysis of the $CoP\text{-}CoG_v$ displacements showed (1) larger RMS in the Biofeedback than No-biofeedback condition along both the ML and AP axes (T=2.22, $P<0.05$, Figure 1A and T=4.96, $P<0.001$, Figure 1B, respectively), and (2) smaller MPF in the Biofeedback than No-biofeedback condition along both the ML and AP axes (T=2.33, $P<0.05$, Figure 1C and T=3.20, $P<0.01$, Figure 1D, respectively).
    Analysis of the $CoG_h$ displacements showed (1) smaller RMS in the Biofeedback than No-biofeedback condition along both the ML and AP axes (T=4.88, $P<0.001$, Figure 1E and T=3.89, $P<0.01$, Figure 1F, respectively), and (2) smaller MPF in the Biofeedback than No-biofeedback condition along both the ML and AP axes (T=4.48, $P<0.001$, Figure 1G and T=3.66, $P<0.01$, Figure 1H, respectively).

-------------------------------------
Please insert Figure 1 about here
-------------------------------------

Stabilogram-diffusion analysis
*Transition point co-ordinates*

Analysis of the CoP-CoG$_v$ displacements showed (1) decreased time intervals Δt of the transition point in the Biofeedback relative to the No-biofeedback condition along both the ML and AP axes (T=2.16, *P*<0.05, Figure 2A and T=4.82, *P*<0.001, Figure 2B, respectively), and (2) increased mean square distances <Δx$^2$> of the transition point in the Biofeedback relative to the No-biofeedback condition along both the ML and AP axes (T=2.31, *P*<0.05, Figure 2C and T=4.31, *P*<0.001, Figure 2D, respectively).

Analysis of the CoG$_h$ displacements showed (1) decreased time intervals Δt of the transition point in the Biofeedback relative to the No-biofeedback condition along both the ML and AP axes (T=2.16, *P*<0.05, Figure 2G and T=4.82, *P*<0.001, Figure 2H, respectively), and (2) decreased mean square distances <Δx$^2$> of the transition point in the Biofeedback relative to the No-biofeedback condition along both the ML and AP axes (T=2.63, *P*<0.05, Figure 2I and T=2.17, *P*<0.05, Figure 2J, respectively).

-------------------------------------
Please insert Figure 2 about here
-------------------------------------

*Scaling regime*s exponents

Analysis of the CoP-CoG$_v$ displacements showed larger short latency scaling regimes exponents H$_{sl}$ in the Biofeedback than No-biofeedback condition along both the ML and AP axes (T=2.81, *P*<0.05, Figure 3A and T=3.62, *P*<0.01, Figure 3B, respectively), suggesting an increased persistent behaviour in the short-term region during the shortest time intervals in the Biofeedback relative to the No-biofeedback condition.

Analysis of the CoG$_h$ displacements showed smaller long latency scaling regimes exponents H$_{ll}$ in the Biofeedback than No-biofeedback condition along both the ML and AP axes (T=7.65, *P*<0.001, Figure 3G and T=4.28, *P*<0.001, Figure 3H, respectively), suggesting an increased anti-persistent behaviour in the long-term region during the longest time intervals in the Biofeedback relative to the No-biofeedback condition.

Finally, as generally observed in this kind of investigation, for both experimental conditions, the results of long latency scaling regimes exponents H$_{ll}$ for CoP-CoG$_v$ displacements (Figures 3C,3D) and those of short latency scaling regimes exponents H$_{sl}$ for CoG$_h$ displacements (Figures 3E,3F) were close to 0.5, hence indicating a behaviour solely stochastic in nature (e.g., Berger et al. 2005; Bernard-Demanze et al. 2006; Rougier et al. 2001; Rougier and Caron 2000; Rougier and Farenc 2000).

-------------------------------------
Please insert Figure 3 about here
-------------------------------------

**Discussion**

The purpose of this study was to determine the effects of a plantar pressure-based, tongue-placed tactile biofeedback on postural control mechanisms during quiet standing. To this aim, sixteen young healthy adults were asked to stand as immobile as possible with their eyes closed in two conditions of No-biofeedback and Biofeedback. Centre of foot pressure (CoP) displacements, recorded using a force platform, were used to compute the horizontal displacements of the vertical projection the centre of gravity (CoG$_h$) and those of the difference between the CoP and the vertical projection of the CoG (CoP-CoG$_v$). These displacements were processed through frequency-domain and stabilogram-diffusion analyses to assess their spatio-temporal linkage and their degree of control.

Effect of biofeedback on CoP-CoG$_v$ displacements

Analysis of the CoP-CoG$_v$ displacements showed increased RMS (Figures 1A,1B) and MPF (Figures 1C,1D) along both the ML and AP axes in the Biofeedback and No-biofeedback condition. Complementary to the frequency-domain analysis, modelling the CoP-CoG$_v$ displacements as fBm through the stabilogram diffusion analysis provided additional insight into the nature and the temporal organisation of the control mechanisms involving the CoP-CoG$_v$ displacements called into play in the Biofeedback condition.

On the one hand, indeed, the increased CoP-CoG$_v$ RMS (Figures 1A,1B) observed in the Biofeedback condition were likely to be related to (1) spatial parameters, since increased spatial co-ordinates of the transition point ($<\Delta x^2>$) were observed along both the ML and AP axes in the Biofeedback relative to the No-biofeedback condition (Figures 2C,2D), and (2) an increased persistent behaviour of CoP-CoG$_v$ displacements in the short-term region during the shortest time intervals, since increased short latency scaling exponents H$_{sl}$ were observed along both the ML and AP axes in the Biofeedback relative to the No-biofeedback condition (Figures 3A,3B).

On the other hand, the increased CoP-CoG$_v$ MPF observed in the Biofeedback condition (Figures 1C,1D) means, by definition, that a diminution of the period needed for the CoP-CoG$_v$ to return to a similar position occurred. This result could hence be related to temporal parameters, since reduced temporal co-ordinates of the transition point $\Delta t$ were observed along both the ML and AP axes in the Biofeedback relative to the No-biofeedback condition (Figures 2A,2B).

Finally, it is important to keep in mind that, from a biomechanical point of view, increasing the amplitudes of CoP-CoG$_v$ displacements in the Biofeedback condition, seen as an expression of the initial horizontal acceleration communicated to the CoG$_h$ (Brenière et al. 1987; Gage et al. 2004; Winter et al. 1998), would negatively affect the relative facility for the subjects to handle CoG$_h$ displacements in this condition due to the lower forces they would have to counteract.

Effect of biofeedback on CoG$_h$ displacements

Analysis of the CoG$_h$ displacements showed decreased RMS (Figures 1E,1F) and increased MPF (Figures 1C,1D) along both the ML and AP axes in the Biofeedback relative to the No-biofeedback condition. At this point, considering the increased CoP-CoG$_v$ amplitudes observed in the Biofeedback relative to the No-biofeedback condition (Figures 1A,1B), determining larger initial horizontal accelerations communicated to the CoG$_h$ (Brenière et al. 1987; Gage et al. 2004; Winter et al. 1998), it was hypothesized these decreased CoG$_h$ amplitudes observed with Biofeedback (Figures 1E,1F) to stem from a modification of the control characteristics of the CoG$_h$ displacements set by the central nervous system, involving (1) a reduction in the correction thresholds and/or (2) an increased efficiency of the corrective mechanisms. Accordingly, modelling the CoG$_h$ displacements as fractional Brownian motion through the stabilogram diffusion analysis allowed providing additional insight into (1) the spatio-temporal coordinates of the transition point at which the corrective process CoG$_h$ displacements is initiated ($<\Delta x^2>$ and $\Delta t$) and (2) the extent to which the CoG$_h$ is controlled (H$_{ll}$).

On the one hand, results showed reduced spatio-temporal co-ordinates of the transition point ($<\Delta x^2>$ and $\Delta t$) along both the ML and AP axes in the Biofeedback relative to the No-biofeedback condition (Figures 2E,2F,2G,2H), suggesting that the distance covered by the CoG$_h$ and the time spent before the onset of a corrective process were reduced in the Biofeedback relative to the No-biofeedback condition.

On the other hand, results showed decreased long latency scaling exponents (H$_{ll}$) in the Biofeedback relative to the No-biofeedback condition (Figures 3G,3H), suggesting an increased probability that CoG$_h$ away from a relative equilibrium point will be offset by corrective adjustments back towards the equilibrium position, once a threshold in sway has been reached, with Biofeedback.

These shorter distances ($<\Delta x^2>$) associated with shorter time intervals ($\Delta t$) before a corrective mechanism begin to operate, and the improved determinism in this corrective process aimed at returning the $CoG_h$ to its initial position ($H_{ll}$) observed along both the ML and AP directions in the Biofeedback relative to the No-biofeedback condition, could be explained by the specificity of the biofeedback provided to the subjects. Indeed, as above-mentioned, the plantar pressure-based, tongue-placed tactile biofeedback used in this study presented the particularity of supplying the subjects with (1) supplementary sway-related cues that could have allowed them to decrease spatio-temporal thresholds from which the postural corrections were set (Figures 2E,2F,2G,2H), and (2) a constant reference position ("dead zone") in which they were required to stay (or, at least return regularly), that could have allowed them to increase the degree to which $CoG_h$ displacements were controlled (Figures 3G,3H). Interestingly, with regard to the provision of enhanced sensory information from the plantar soles to the postural control system, our results replicate those of two recent studies investigating, with the same analysis method, the effects of an increased sensitivity of the plantar mechanoreceptors on postural control mechanisms during quiet standing (Bernard-Demanze et al. 2004,2006). Indeed, plantar soles massages have been shown to induce, along both the ML and AP axes, reduced spatio-temporal co-ordinates of the transition point at which the corrective process involving $CoG_h$ displacements is initiated (as indicated by decreased mean square distances $<\Delta x^2>$ and time intervals $\Delta t$) and an improved determinism in this corrective process aimed at returning the $CoG_h$ to an equilibrium point (as indicated by decreased diminished long latency scaling exponents $H_{ll}$). What is more, the magnitude of these effects on $CoG_h$ displacements was reported to increase with increasing the duration of the plantar soles massage (Bernard-Demanze et al. 2006). Along theses lines, it is possible that modifying the size of the predetermined adjustable "dead zone" of the biofeedback also would affect postural control mechanisms during quiet standing observed in the present study. An experiment is currently being performed to address this issue.

**Conclusion**

In light of the CoP displacements dissociation into two elementary $CoP$-$CoG_v$ and $CoG_h$ displacements and the recourse to fBm modelling through the stabilogram-diffusion analysis, the present findings suggest that the main way the plantar pressure-based, tongue-placed tactile biofeedback improves postural control during quiet standing (Vuillerme et al. 2006a) is via both a reduction of the correction thresholds and an increased efficiency of the corrective mechanism involving $CoG_h$ displacements.

**Acknowledgements**

The authors are indebted to Professor Paul Bach-y-Rita for introducing us to the Tongue Display Unit and for discussions about sensory substitution. The authors would like to thank subject volunteers. The company Vista Medical is acknowledged for supplying the FSA Inshoe Foot pressure mapping system. This research was supported by the company IDS, Floralis (Université Joseph Fourier, Grenoble), and GRenoble Alpes Valorisation Innovation Technologies.

**References**

Bach-y-Rita P, Kaczmarek KA, Tyler ME, Garcia-Lara J (1998) Form perception with a 49-point electrotactile stimulus array on the tongue. J Rehabil Res Dev 35:427-430.
Berger L, Chuzel M, Buisson G, Rougier P (2005) Undisturbed stance control in the elderly: Part 1. Age-related chages undisturbed upright stance control. J Mot Behav 37:348-358.
Bernard-Demanze L, Burdet C, Berger L, Rougier P (2004) Recalibration of somesthetic plantar information in the control of undisturbed upright stance maintenance. J Integr Neurosci 3:433-451.
Bernard-Demanze L, Vuillerme N, Berger L, Rougier P (2006) Magnitude and duration of the effects of plantar sole massages. International SportMed Journal. 7:154-169.
Brenière Y (1996) Why do we walk the way we do? J Mot Behav 28:291-298.
Brenière Y, Do MC, Bouisset S (1987) Are dynamic phenomena prior to stepping essential to walking? J Mot Behav 19:62-76.
Caron O, Faure B, Brenière Y (1997) Estimating the center of gravity of the body on the basis of the center of pressure in standing posture. J Biomech 30:1169-1171.
Caron O, Gelat T, Rougier P, Blanchi JP (2000) A comparative analysis of the centre of gravity and centre of pressure sway paths in standing posture: an estimation of active stiffness. J Appl Biomech 16:234-247.
Collins JJ, De Luca CJ (1993) Open-loop and closed-loop control of posture: a random-walk analysis of centre-of-pressure trajectories. Exp Brain Res 95:308-318.
Clément G, Gurfinkel VS, Lestienne F, Lipshits MI, Popov KE (1984) Adaptation of postural control to weightlessness. Exp Brain Res 57:61-72.
Essick GK, Chopra A, Guest S, McGlone F (2003) Lingual tactile acuity, taste perception, and the density and diameter of fungiform papillae in female subjects. Physiol Behav 80:289-302.
Gage WH, Winter DA, Frank JS, Adkin AL (2004) Kinematic and kinetic validity of the inverted pendulum model in quiet standing. Gait & Posture 19:124-132.
Horstmann A, Dietz V (1990) A basic posture control mechanism: the stabilization of the centre of gravity. Electroenceph Clin Neurophysiol 76:165-176.
Mandelbrot BB, van Ness JW (1968) Fractional Brownian displacements, fractional noises and applications. SIAM Review 10:422-437.
Rougier P, Burdet C, Farenc I, Berger L (2001) How leaning backward or forward affects undisturbed stance control in humans. Neurosci Res 41:41-50.
Rougier P, Caron O (2000) Centre of gravity displacements and ankle joint stiffness control in upright undisturbed stance modeled through a fractional Brownian motion framework. J Mot Behav 32:405-413.
Rougier P, Farenc I. (2000) Adaptative effects of loss of vision on upright undisturbed stance. Brain Res 871:165-174.
Vuillerme N, Chenu O, Demongeot J, Payan Y (2006a) Controlling posture using a plantar pressure-based, tongue-placed tactile biofeedback system. Exp. Brain Res. (DOI 10.1007/s00221-006-0800-4).
Vuillerme N, Chenu O, Demongeot J, Payan Y (2006b) Improving human ankle joint position sense using an artificial tongue-placed tactile biofeedback. Neurosci Lett 405:19-23.
Vuillerme N, Chenu O, Fleury, J, Demongeot J, Payan Y (2006c) Optimizing the use of an artificial tongue-placed tactile biofeedback for improving ankle joint position sense in humans. 28th Annual International Conference of the IEEE Engineering in Medicine and Biology Society (EMBS), New York, USA.
Winter DA, Prince F, Frank JS, Powell C, Zabjek KF (1996) Unified theory regarding A/P and M/L balance in quiet stance. J Neurophysiol 75:2334-2343.

Winter DA, Patla AE, Prince F, Ishac M, Gielo-Perczak K (1998) Stiffness control of balance in quiet standing. J Neurophysiol 80:1211-1221.

**Figure captions**

**Figure 1.** Mean and standard deviation of the root mean square (RMS) along the ML (A,E) and AP (B,F) axes and mean power frequencies (MPF) along the ML (C,G) and AP (D,H) axes for CoP-CoG$_v$ and CoG$_h$ displacements obtained in the No-biofeedback and Biofeedback conditions. These two experimental conditions are presented with different symbols: No-biofeedback (*white bars*) and Biofeedback (*black bars*). Upper and lower panels represent CoP-CoG$_v$ and CoG$_h$ displacements, respectively. The significant *P* values for comparisons between No-biofeedback and Biofeedback conditions also are reported (*: *P*<0.05; **: *P*<0.01; ***: *P*<0.001).

**Figure 2.** Mean and standard deviation of the temporal co-ordinates of the transition point (Δt) along the ML (A,E) and AP axes (B,F), and the spatial co-ordinates (<Δx$^2$>) of the transition point along the ML (C,G) and AP axes (D,H) for CoP-CoG$_v$ and CoG$_h$ displacements obtained in the No-biofeedback and Biofeedback conditions. These two experimental conditions are presented with different symbols: No-biofeedback (*white bars*) and Biofeedback (*black bars*). Upper and lower panels represent CoP-CoG$_v$ and CoG$_h$ displacements, respectively. The significant *P* values for comparisons between No-biofeedback and Biofeedback conditions also are reported (*: *P*<0.05; ***: *P*<0.001).

**Figure 3.** Mean and standard deviation of the short latency scaling exponents (H$_{sl}$) along the ML (A,E) and AP axes (B,F), and the long along latency scaling exponents (H$_{ll}$) along the ML (C,G) and AP axes (D,H) for CoP-CoG$_v$ and CoG$_h$ displacements obtained in the No-biofeedback and Biofeedback conditions. These two experimental conditions are presented with different symbols: No-biofeedback (*white bars*) and Biofeedback (*black bars*). Upper and lower panels represent CoP-CoG$_v$ and CoG$_h$ displacements, respectively. The significant *P* values for comparisons between No-biofeedback and Biofeedback conditions also are reported (*: *P*<0.05; **: *P*<0.01; ***: *P*<0.001).

**Figure 1.**

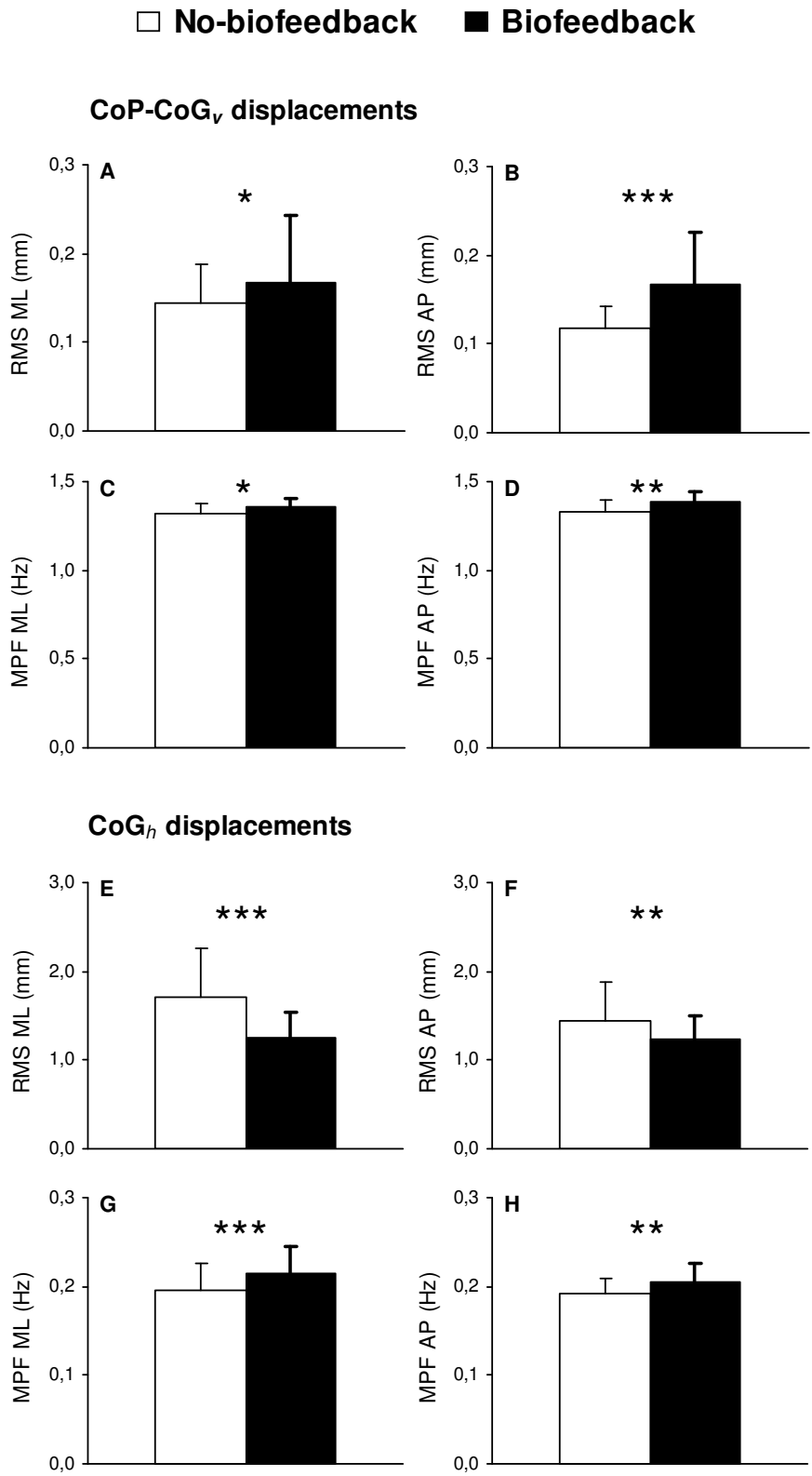

**Figure 2.**

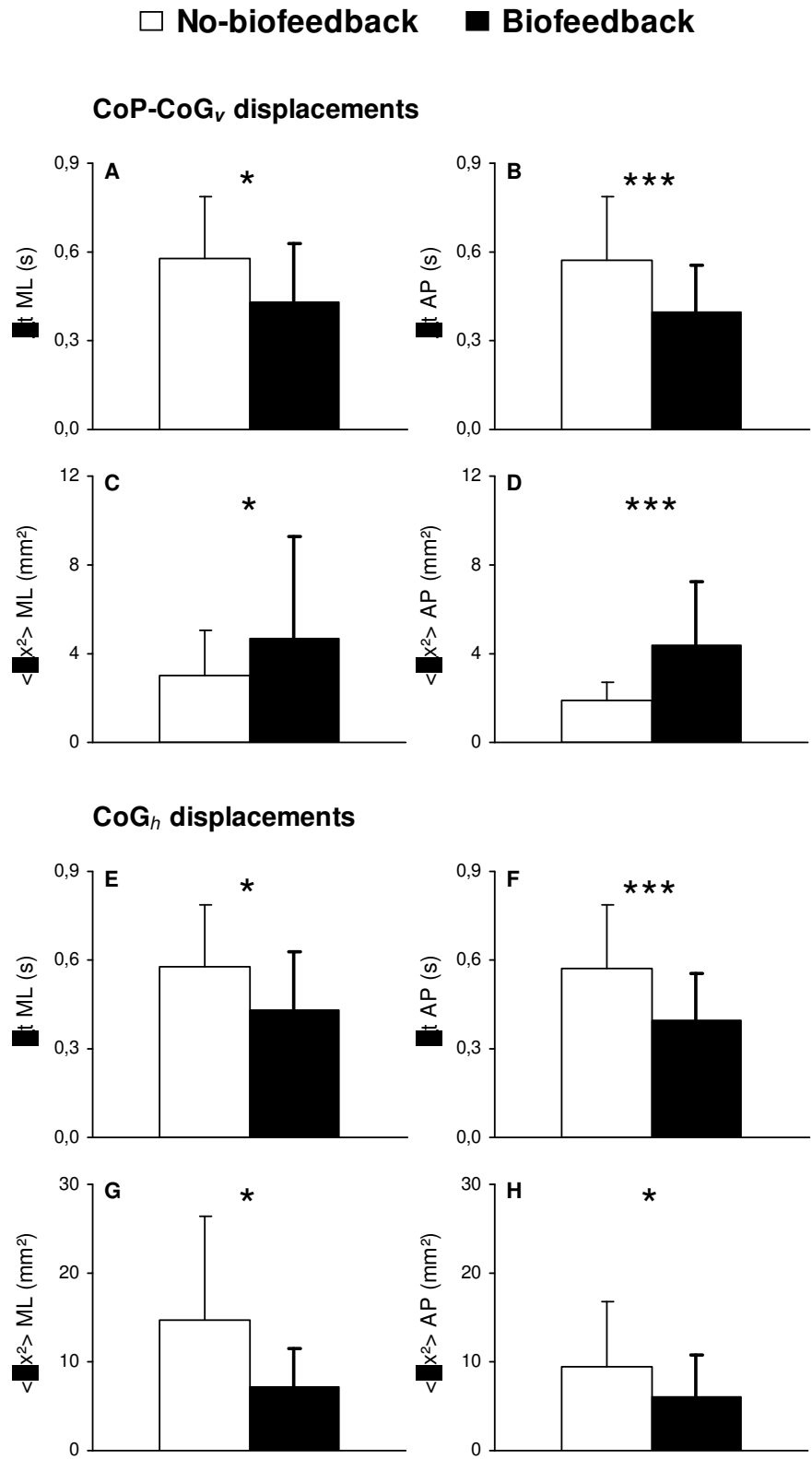

**Figure 3.**

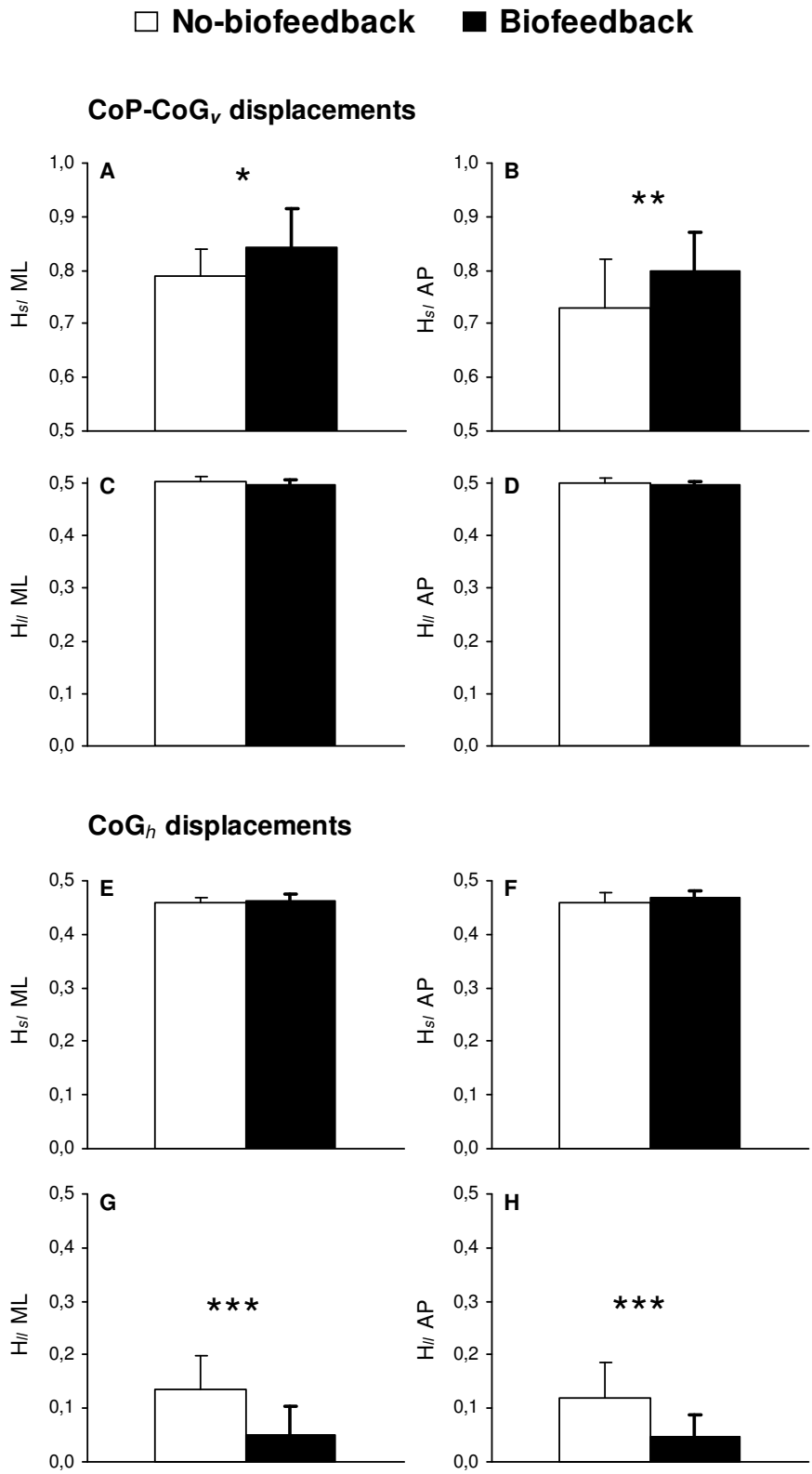